# Role-Playing LLM-Based Multi-Agent Support Framework for Detecting and Addressing Family Communication Bias

Rushia Harada
*Almondays Co., Ltd*
*School of Economics and Business Administration*
*Yokohama City University,*
*Yokohama, Japan*
*Email: c221219f@yokohama-cu.ac.jp*

Yuken Kimura
*Almondays Co., Ltd*
*School of Economics and Business Administration*
*Yokohama City University,*
*Yokohama, Japan*
*Email: s222045f@yokohama-cu.ac.jp*

Keito Inoshita
*Almondays Co., Ltd*
*Faculty of Business and Commerce*
*Kansai University,*
*Osaka, Japan*
*Email: inosita.2865@gmail.com*

***Abstract*—Well-being in family settings involves subtle psychological dynamics that conventional metrics often overlook. In particular, unconscious parental expectations, termed ideal parent bias, can suppress children's emotional expression and autonomy. This suppression, referred to as suppressed emotion, often stems from well-meaning but value-driven communication, which is hard to detect or address from outside the family context. Focusing on these latent dynamics, this study explores Large Language Model (LLM)-based support for psychologically safe family communication. We constructed a Japanese parent–child dialogue corpus of 30 scenarios, each annotated with metadata on ideal parent bias and suppressed emotion. Based on this corpus, we developed a role-playing LLM-based multi-agent dialogue support framework that analyzes dialogue and generates feedback. Specialized agents detect suppressed emotion, describe implicit ideal parent bias in parental speech, and infer contextual attributes such as the child's age and background. A meta-agent compiles these outputs into a structured report, which is then passed to five selected expert agents. These agents collaboratively generate empathetic and actionable feedback through a structured four-step discussion process. Experiments show that the system can detect categories of suppressed emotion with reasonable discrimination, and produce feedback rated highly in empathy and practicality. Moreover, simulated follow-up dialogues incorporating this feedback exhibited signs of improved emotional expression and mutual understanding, suggesting the framework's potential to support positive transformation in family interactions.**

***Keywords—Suppressed Emotion, Ideal Parent Bias, Multi-Agent System, Communication Support, Large Language Model***

## I. INTRODUCTION

Well-being is a multidimensional concept that covers not only objective indicators such as economic growth or health status, but also subjective satisfaction and the quality of social relationships [1]. Organizations such as the World Health Organization (WHO) and Organisation for Economic Co-operation and Development (OECD) have advocated for incorporating psychosocial indicators, such as life satisfaction and social trust, into policy evaluation, which has led to increased attention on technologies that support psychological well-being through everyday communication, especially those focusing on the quality of dialogue [2]. The family, as the micro-environment where individuals spend most of their time, plays a critical role in developing children's emotional regulation, self-esteem, and coping skills [3]. In practice, however, unconscious parental expectations, such as academic superiority or traditional gender roles, often appear in communication and unintentionally suppress children's freedom of expression and behavioral exploration. This study refers to such expectations as ideal parent bias. Children may also experience anxiety or frustration in response to these interactions but may find themselves unable to express such emotions openly. This internalization is defined as suppressed emotion, which can adversely affect mental health and social relationships later in life. Since these forms of bias and suppression are typically rooted in goodwill and lack overt dominance, they are difficult to recognize or intervene in from outside the family.

This study focuses on the interaction between ideal parent bias and suppressed emotion and aims to develop a role-playing Large Language Model (LLM)-based multi-agent dialogue support framework that facilitates transformation and repair in parent–child relationships. To this end, we construct a set of realistic Japanese parent–child dialogue scenarios involving both phenomena, developed through collaboration between LLMs and human annotators. Using this data and a multi-agent LLM system, we explore the following research questions. First, we examine whether the framework can accurately detect suppressed emotion and ideal parent bias and explain them in a contextually grounded and human-interpretable manner. Second, we assess whether the framework can generate feedback that is both empathetic and actionable by integrating diverse perspectives from role-played expert agents and tailoring the output to each family member. Third, we evaluate whether such feedback can contribute to desirable changes in dialogue or behavior, as reflected in simulated post-feedback conversations, including reduced suppression, reduced bias, and enhanced self-expression. All evaluations prioritize human intuition and subjective plausibility, aiming to support emotionally resonant LLM dialogue assistance beyond mere output accuracy.

This study makes three main contributions:

i) Construction of a Japanese parent–child dialogue corpus containing 30 scenarios annotated with suppressed emotion and ideal parent bias, enabling empirical investigation into relational repair through dialogue.
ii) Development of a multi-agent framework that detects suppressed emotion and ideal parent bias, estimates relevant attributes (e.g., age, background), and generates empathetic and practical feedback through collaborative expert discussion.
iii) Proposal of an evaluation method that simulates post-feedback dialogue using agent-based role-play, allowing multifaceted assessment of behavioral and relational transformation.

The rest of this paper is organized as follows. Section II reviews related work and technical background. Section III describes the construction of the dialogue dataset. Section IV presents the architecture and design of our proposed system.

Section V reports experimental results. Section VI concludes the paper and outlines future directions.

## II. Related Work

### A. Parental Value Bias and Its Impact on Children

Parental values and family communication styles significantly influence children's emotional development and identity formation. Recent studies have examined the multifaceted impact of unconscious parenting attitudes on children. Odintsova et al. demonstrated that emotional suppression and overcontrol by parents are associated with reduced resilience in adulthood. Amelia et al. [4] found that parental communication ability has a notable effect on children's social and emotional development. Son and Kim [5] reported that open communication during difficult situations contributes to children's psychological stability and adaptability. Blossom et al. [6] showed that parental anticipation of threats and family dysfunction are linked to distorted threat perception in children. Rudnova et al. [7] argued that a parent's belief in structuring the family environment contributes to cognitive development. Nasir and Johari [8] pointed out that conversation-oriented households tend to promote emotional regulation in children, while conformity-oriented ones are more likely to intensify suppression. These studies collectively highlight how parental values and communication styles shape children's suppressed emotions and psychological adjustment, reinforcing the relevance and necessity of the framework proposed in this study.

### B. Multi-Agent LLM for Psychological Support

Recent developments in multi-agent dialogue systems using LLMs have attracted attention in fields such as psychological support and bias detection. Xie et al. [9] introduced MindScope, a framework for identifying 72 types of cognitive biases through inter-agent discussion. Wu et al. [10] explored the generation of empathetic responses by simulating discussions among agents modeled after different psychological schools. Kampman et al. [11] showed that multiple agents can collaboratively assist clinical professionals through response generation and summarization. Ke et al. [12] demonstrated that multi-agent debate helps correct cognitive biases and significantly improves diagnostic accuracy in medical contexts. Chang [13] proposed SocraSynth, a system for consensus-building via cooperative argumentation, and introduced a framework for adversarial dialogue aimed at bias detection. Chen and Liu [14] developed MADP, a support model based on cognitive behavioral therapy, in which multiple agents share responsibilities for emotional and cognitive processing. These works demonstrate the practicality and flexibility of LLM-based multi-agent dialogue in tasks such as bias detection and empathetic support, forming a technological foundation for the present study.

## III. Dataset Construction Methodology

This study constructs a Japanese dialogue dataset for detecting and intervening in suppressed emotion and ideal parent bias in parent–child interactions. The 30 scenarios use predefined parent–child personas varying in gender, age (6–15), personality, and family background. Child attributes include shyness, extroversion, or resistance to authority, while parent personas reflect biases such as academic pressure, gender norms, role-based expectations, and self-esteem projection, defined together with upbringing history. For example, one scenario features a father from a “sports-elite” family who believes “non-athletic children are inferior,” interacting with his shy 9-year-old child.

Each scenario includes metadata that captures both the parent’s bias background and the child’s emotional suppression tendencies. These annotations include the type and strength of the bias, the underlying background of its formation, the presence or absence of suppression, the type and strength of suppression, among others. Each scenario also features a specific topic, such as play, study, or future plans, and follows a consistent structure of 10 dialogue turns, from introduction and conflict to resolution. This structure allows for clear observation of emotional shifts and stylistic changes in language.

Based on this design, dialogue generation was conducted using three types of agents:

- Parent agent: generates value-driven utterances (e.g., “you should...”) according to the defined type and intensity of ideal parent bias.
- Child agent: produces responses based on the child’s age, personality, and suppression level, choosing from behaviors such as compliance, resistance, or silence.
- Narrator agent: adds psychological annotations and contextual descriptions during the dialogue to supplement background understanding.

Each dialogue was generated turn-by-turn in a role-play format, with prompts specifying the topic, roles, and stopping conditions. Outputs included utterance number, speaker, and content, stored in CSV format. To ensure quality, experts in parent–child communication refined linguistic style and age-appropriate expressions. For instance, stronger parental bias was expressed with more commands or negations, while child utterances were simplified into short, age-suitable phrases.

As a result, the dataset was constructed as a high-fidelity, human-reviewed parent–child dialogue resource in which both utterances and contextual structures are controllable. It provides a consistent foundation for evaluating the detection, intervention, and behavioral transformation of suppressed emotion and ideal parent bias.

## IV. Methodology

### A. Framework Overview

Fig. 1 illustrates the overall architecture of the system. The proposed framework detects suppressed emotion and ideal parent bias in parent–child dialogue and generates feedback that is both empathetic and actionable. Given a dialogue input $D$, the suppressed emotion detection agent $A_{sup}$ identifies the child’s suppressed emotion, while the attribute completion agent $A_{attr}$ estimates auxiliary attributes such as the child’s gender, age, and family background. The bias detection agent $A_{bias}$ then generates a description of the ideal parent bias latent in the parent’s speech. A meta-agent $A_{meta}$ integrates the outputs from these agents into a situation report $R$, which serves as the basis for generating feedback. The feedback is produced by a selected set of five expert agents $E_{select} = \{E_1, E_2, \ldots, E_5\}$, chosen based on embedding similarity and viewpoint diversity. Each expert agent reviews the report $R$ and the dialogue $D$, provides an initial opinion, and comments on the feedback proposed by others. A final meta-agent $A_{final}$ synthesizes these perspectives into a unified and nuanced response tailored to the dialogue participants.

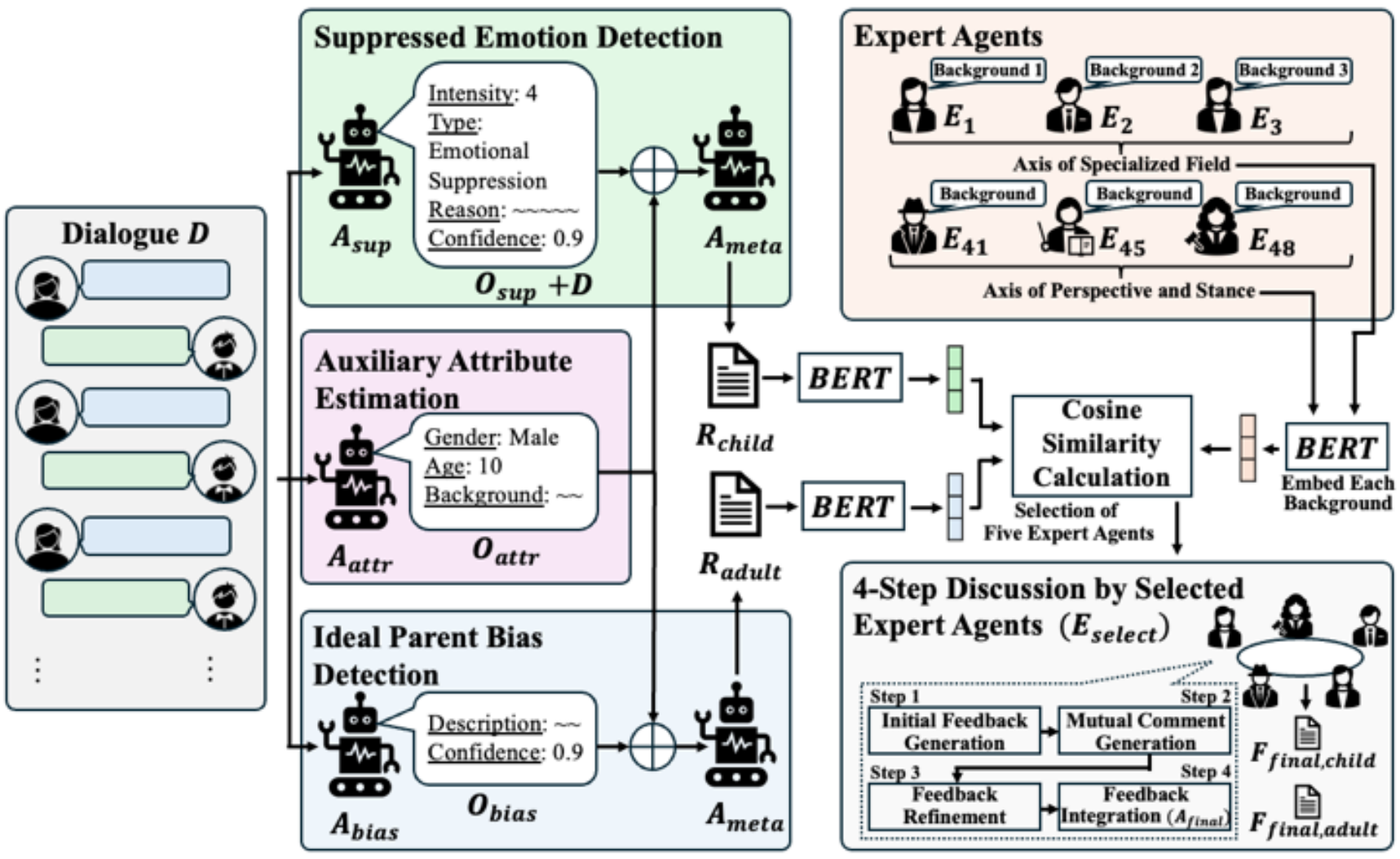

Fig. 1. Framework Overview of the Role-Playing LLM-Based Multi-Agent Dialogue Support.

In this way, the framework leverages LLMs in a multi-stage, multi-agent structure to generate feedback $F_{final}$ that supports emotionally attuned communication, promotes psychological safety, and fosters the repair of strained relationships in parent–child interactions.

*B. A Multi-Stage Suppressed Emotion Detection*

To detect suppressed emotion in parent–child dialogue, this study introduces a multi-stage suppressed emotion detection framework based on LLMs. The system takes a dialogue input $D$ and performs two parallel processes: detection of suppressed emotion and estimation of auxiliary attributes. These outputs are then integrated into a report by a meta-agent.

The suppressed emotion detection agent $A_{sup}$, designed to emulate the role of a psychological counselor, takes dialogue $D$ as input and produces a four-element output:

$$A_{sup}(D) \rightarrow O_{sup} = \{s, suppression_{type}, reason, c\} \quad (1)$$

In this output, the variable $s \in [1, 5]$ represents the intensity of suppression on a five-point scale. The field $suppression_type \in S$ denotes the category of the emotion being suppressed, where the category set $S$ consists of six predefined types: anxiety, emotional, fear, social, behavioral, and self-esteem. The value of $reason$ is a natural language explanation generated by the agent to clarify the psychological background of the suppression. The confidence score $c \in [0.0, 1.0]$ quantifies the model's certainty in the detection result.

To complement this detection, the attribute estimation agent $A_{attr}$ analyzes the same dialogue $D$ and returns auxiliary information:

$$A_{attr}(D) \rightarrow O_{attr} = \{gender, age, background\} \quad (2)$$

In this case, $gender$ indicates the child's gender as either male or female, while $a \in \mathbb{Z}_+$ denotes the child's estimated age in years. The variable $background$ provides a brief natural language summary of the child's family context.

These two outputs, $O_{sup}$ and $O_{attr}$, are then passed to a meta-agent $A_{meta}$, which synthesizes them into a structured situation report:

$$A_{meta}(D, O_{sup}, O_{attr}) \rightarrow R_{child} \quad (3)$$

The resulting report $R_{child}$ offers a coherent summary of the child's emotional state and personal background. It is used by expert agents in the feedback generation phase to deliver context-sensitive and psychologically appropriate responses.

*C. A Multi-Stage Ideal Parent Bias Interpretation*

To visualize unconscious ideal parent bias embedded in parental speech, the system introduces an agent that extracts descriptive features from dialogue using natural language. Given dialogue input $D$, the bias detection agent $A_{bias}$ generates a description of implicit value-based expectations, and this output is combined with auxiliary attributes $O_{attr}$. The meta-agent $A_{meta}$ then integrates these into a structured report for further analysis.

The process begins with agent $A_{bias}$ receiving the dialogue $D$ and producing the following output:

$$A_{bias}(D) \rightarrow O_{bias} = \{bias\ description, c\} \quad (4)$$

Here, $bias\ description$ is a natural language explanation of the ideal parent bias embedded in the parent's speech, and $c \in [0.0, 1.0]$ is the confidence score assigned by the LLM.

To support this task, the system provides example categories of ideal parent bias through prompts. However, the model is also encouraged to output hybrid or nuanced descriptions, allowing for flexible expression that better reflects real-world complexity. This approach helps generate feedback that captures subtle interpersonal dynamics and while avoiding overly rigid classifications. The predefined categories used in prompt design are summarized in the Table I.

TABLE I. CATEGORIES OF IDEAL PARENT BIAS USED IN PROMPTS

| Category | Example Expressions |
|---|---|
| Academic Excellence | "You have to get good grades" (academic pressure on performance) |
| Gender Norms | "Boys shouldn't cry" (traditional gender role expectations) |
| Social Comparison | "Why can't you be like your sibling?" (comparison with peers or siblings) |
| Self-replication | "When I was your age..." (projecting the parent's own values and experiences) |
| Role-based Expectation | "As a big sister, you should…" (expectations based on family roles) |
| Self-esteem Projection | "You're embarrassing me," (concerns about parental status or social image) |

Following bias detection, the system combines the bias output $O_{bias}$ with the previously estimated attributes $O_{attr}$. The meta-agent $A_{meta}$ then consolidates these into a situation report:

$$A_{meta}(D, O_{bias}, O_{attr}) \rightarrow R_{adult} \tag{5}$$

This report $R_{adult}$ organizes the parent's underlying values and contextual background and is used in the subsequent feedback generation phase. The format and structure mirror the child-side report $R_{child}$, ensuring coherence in analysis across family members.

### D. Role-Playing LLM-Based Multi-Agent Feedback

This section describes the framework for generating empathetic and actionable feedback, based on the detection reports $R_{child}$ and $R_{adult}$, which represent suppressed emotion and ideal parent bias in the dialogue. The feedback framework consists of three parts: agent design, agent selection, and feedback generation through a structured four-step discussion process.

To ensure a wide range of perspectives, we designed a pool of 50 role-playing expert agents $E = \{E_1, E_2, \ldots, E_{50}\}$, each defined along two conceptual axes. The first axis is domain expertise, covering eight fields relevant to child–parent communication, such as psychology, education, social work, and caregiving, with five agents per domain (totaling 40). The second axis captures perspective and stance, including roles such as lived-experience parents, nonviolent communication facilitators, paternal caregivers, and philosophers, ensuring diverse value orientations beyond disciplinary lines.

Each agent $E_i$ is associated with a background document $B_i$, describing their professional knowledge and stance. We encoded all $B_i$ into 768-dimensional vectors using a pre-trained BERT model [15]. Similarly, we embedded the pair of dialogue analysis reports $R_{child,i}$ and $R_{adult,i}$ for each scenario. We calculated cosine similarity between each $B_i$, and the report pair to determine semantic closeness. Based on this score, we automatically selected the top three agents $\{E_1, E_2, E_3\}$ for each scenario. To further increase viewpoint diversity, we used an LLM to select two additional agents $E_4$ and $E_5$ from the lower half of the similarity ranking. These agents were selected to represent differing viewpoints from those of the top three and to enrich the discussion. As a result, we obtained a scenario-specific subset of five expert agents $E_{select} = \{E_1, E_2, \ldots, E_5\}$.

We then simulated collaborative decision-making among these five role-playing expert agents using the following four-step process.

- Step 1: Initial Feedback Generation

Each agent $E_i \in E_{select}$ receives the dialogue $D$ and the combined report $R$ as input and outputs an initial feedback draft:

$$E_i(D, R) \rightarrow \left\{F_i^{(0)}\right\} \tag{6}$$

- Step 2: Peer Commentary Generation

Each agent $E_i$ reviews the initial feedback from other agents $E_j (j \neq i)$ and provides peer comments:

$$E_i\left(F_i^{(0)}\right) \rightarrow \{C_{j\rightarrow i}\} \tag{7}$$

Each agent collects 4 comments:

$$C_i = \{C_{j\rightarrow i} | j \neq i\} \tag{8}$$

- Step 3: Feedback Refinement

Each agent refines its feedback by incorporating received comments $C_i$, producing a revised version:

$$E_i\left(F_i^{(0)}, C_i\right) \rightarrow \left\{F_i^{(1)}\right\} \tag{9}$$

- Step 4: Meta-Integration of Final Feedback

A meta-agent $A_{final}$ synthesizes the refined feedbacks from all five agents $F^{(1)} = \left\{F_1^{(1)}, F_2^{(1)}, \ldots, F_5^{(1)}\right\}$:

$$A_{final}\left(F^{(1)}\right) \rightarrow \{F_{final}\} \tag{10}$$

This role-playing multi-agent mechanism enables feedback generation that is both multifaceted and context-sensitive, offering psychologically safe, constructive suggestions for each family member.

## V. Experiment and Analysis

### A. Experiment Setup

We evaluated the proposed framework on the constructed dataset of 30 Japanese parent–child dialogues $D$, each annotated with instances of suppressed emotion and ideal parent bias. Three raters in their 20s from Almondays Inc., with no specialized training in psychology, conducted the evaluations. Their assessments were grounded in intuitive and empathetic judgments on family communication. We used three criteria: (1) detection accuracy for suppressed emotion types and the appropriateness of ideal parent bias descriptions; (2) relevance and practicality of the generated feedback; and (3) potential for behavioral change induced by the feedback. Raters scored each item using a five-point Likert scale and subjective confidence ratings. We compared these with the framework's outputs to quantify the alignment between machine reasoning and human perception. All LLM-based agents in the framework were implemented using OpenAI's GPT-4o-mini API, with all hyperparameters set to default values.

### B. Evaluation on a Multi-Stage Detection for Suppression and Bias

We evaluated RQ1: whether the multi-stage suppressed emotion detection framework can accurately detect suppressed emotion and ideal parent bias in a human-interpretable manner. Two axes were examined: the suppressed emotion labels assigned to child utterances and the bias descriptions inferred from parent utterances.

For suppressed emotion classification, we compared our framework-predicted labels with majority-vote human annotations. As shown in Table II, the system achieved an accuracy of 0.433, a precision of 0.620, and an F1-score of 0.469. These results demonstrate that the framework can reasonably discriminate suppressed emotion categories, particularly anxiety suppression, while also revealing challenges in recall and overall coverage. In other words, the system shows promising detection capability but still struggles to consistently capture more subtle forms of suppression.

TABLE II. Performance of Suppressed Emotion Classification

| Metric | Accuracy | Precision | Recall | F1 |
|---|---|---|---|---|
| Value | 0.433 | 0.620 | 0.433 | 0.469 |

The confusion matrix in Fig. 2 shows that anxiety suppression was correctly classified in 5 out of 6 cases,

indicating robustness in this category. In contrast, behavioral suppression was frequently misclassified, often confused with emotional or self-esteem suppression. This suggests that the multi-stage suppressed emotion detection framework struggles when multiple suppression cues co-occur in dialogue, blurring decision boundaries.

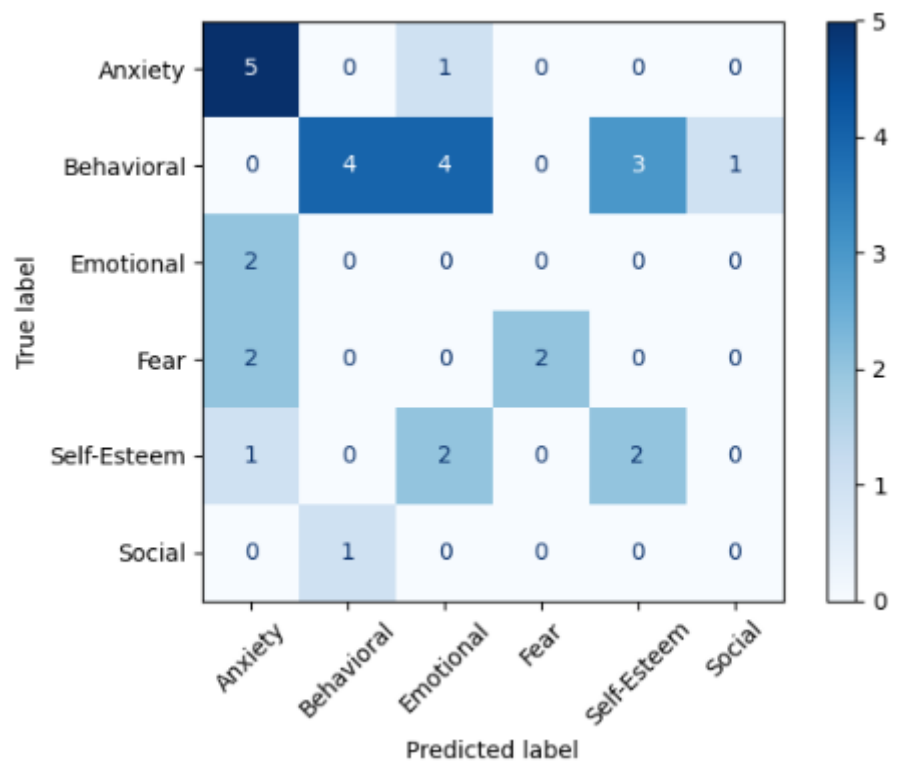

Fig. 2. Confusion Matrix for Suppression Types.

We also analyzed the confidence scores. As illustrated in Fig. 3, the multi-stage suppressed emotion detection framework consistently outputs high confidence levels near 0.85, whereas human raters displayed more moderate levels. This discrepancy suggests that the framework tends to overestimate its certainty, underscoring the need for future calibration to avoid excessive trust in framework's outputs.

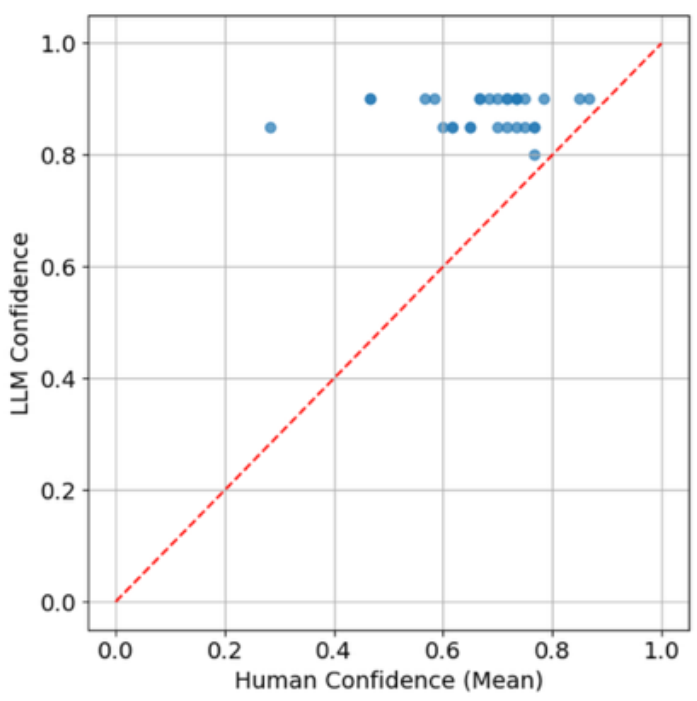

Fig. 3. Our Framework vs. Human Confidence Scores.

For ideal parent bias detection, the three raters showed varying levels of agreement. As shown in Fig. 4, the median score was generally high, suggesting that the bias descriptions were compelling. However, the boxplot for Rater 3 shows greater variance, implying that subjective interpretation and reading context may influence judgment. This variability highlights the inherently ambiguous nature of the "ideal parent bias" concept.

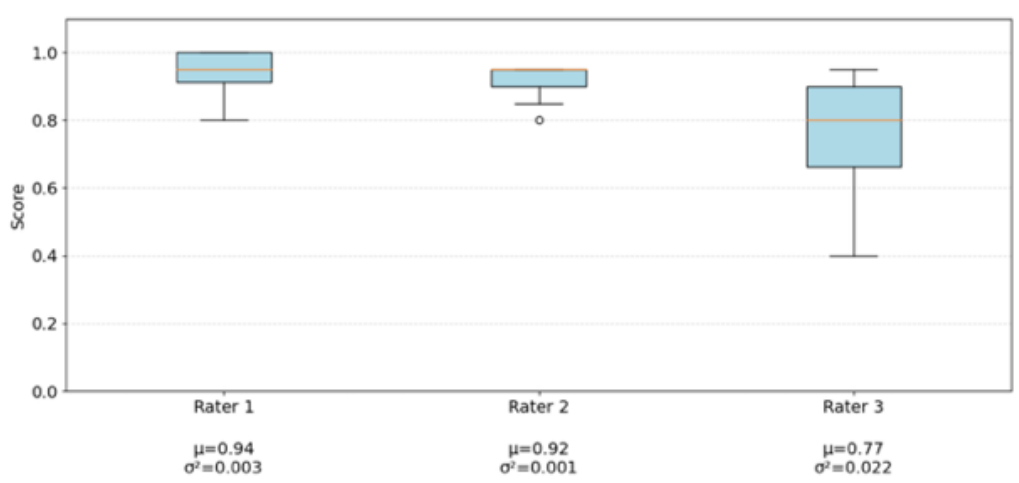

Fig. 4. Human Ratings of Ideal Parent Bias Descriptions.

Finally, attribute estimation, particularly for age, yielded strong performance. The mean absolute error for age prediction was 1.97, suggesting that the attribute completion agent $A_{attr}$ is reasonably effective in estimating background variables. In summary, the system showed moderate performance in detecting specific suppression types and describing implicit bias. However, the overconfidence in outputs and the contextual sensitivity of language generation indicate that our framework may still diverge from human intuition in nuanced scenarios.

### C. Evaluation on Role-Playing LLM-Based Multi-Agent Feedback

We evaluated RQ2: whether the framework can generate feedback that is both empathetic and actionable by integrating diverse perspectives from role-played expert agents and tailoring the output to each family member. To assess this, we conducted a human evaluation of the feedback outputs for both child and parent participants, denoted as $F_{final,child}$ and $F_{final,adult}$. Three raters assessed the outputs using an original 5-point scale based on eight evaluation criteria, separately defined for child and parent feedback.

Fig. 5 shows the results for child-directed feedback. The criteria include: Empathy, Safety, Clarity, Actionability, Self-esteem, Culture, and Developmental fit (Dev_fit). An overall score (Overall) summarizes each rater's impression. The feedback was consistently rated highly for emotional alignment, psychological safety, and linguistic clarity. Rater 1 gave perfect scores in nearly all categories. Rater 3, however, gave slightly lower ratings for Actionability and Self-esteem, suggesting room for improvement in guiding concrete behavior change. These findings indicate that while the framework excels at empathetic language generation and developmental alignment, it may need refinement in proposing tangible actions.

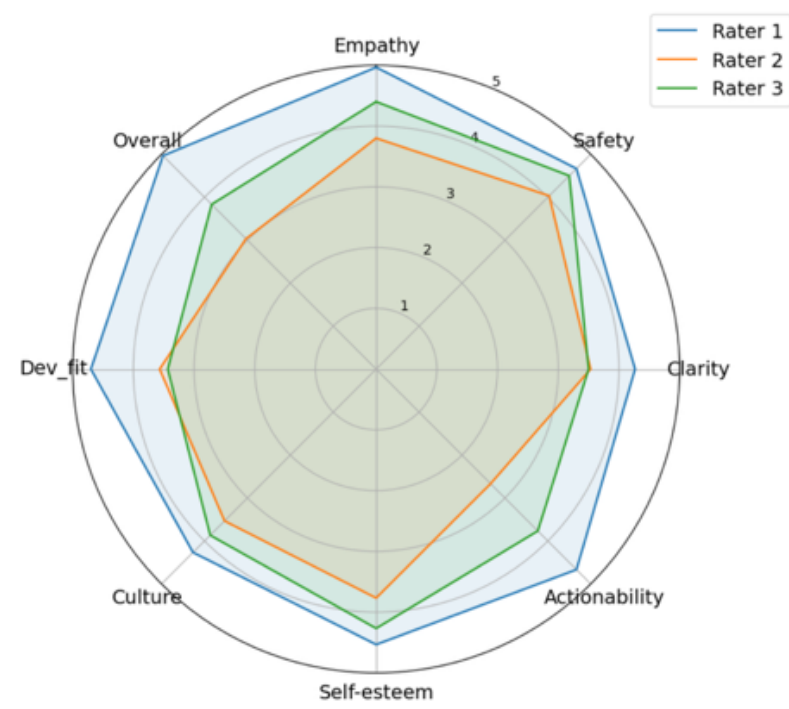

Fig. 5. Human Evaluation of Feedback for Child.

Fig. 6 presents the ratings for parent-directed feedback. Evaluation criteria include: Parent empathy, Non-judgmental tone, Clarity, Actionability, Bias awareness, Child respect, and Context sensitivity. Again, the Overall score aggregates each rater's judgment. Feedback received strong ratings for its respectful tone, awareness-raising of implicit bias, and child-centered framing, particularly from Rater 1 and Rater 3. These raters appeared to appreciate the system's ability to surface latent ideal parent bias in a non-critical manner. By contrast, Rater 2 consistently gave lower ratings, especially for Bias awareness and Clarity, possibly perceiving the phrasing as ambiguous or insufficiently directive. This discrepancy highlights the challenge of crafting universally interpretable responses when addressing sensitive topics such as implicit bias.

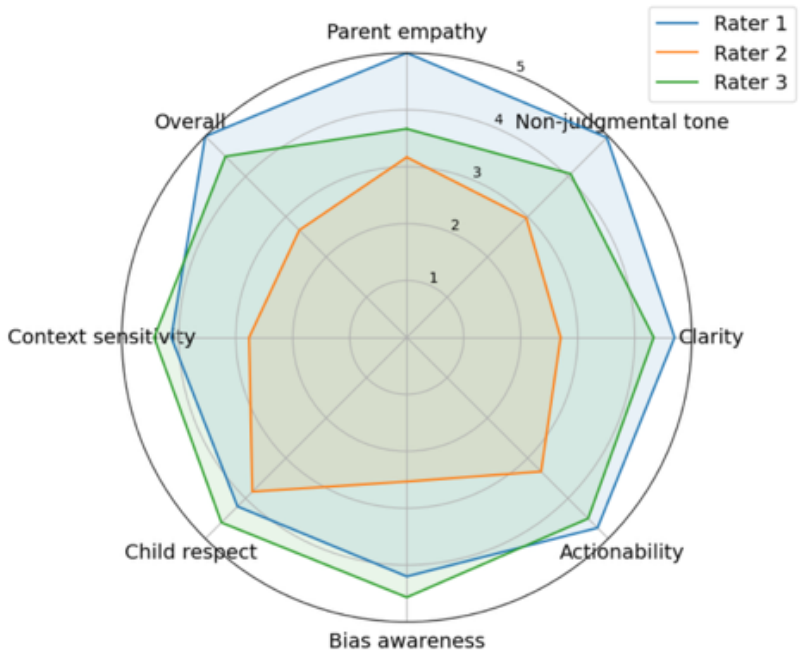

Fig. 6. Human Evaluation of Feedback for Parent.

In summary, the proposed framework was generally well received across both target groups. It demonstrated competence in generating feedback that balances empathy and practical guidance. Nonetheless, the observed variance in human judgments suggests the importance of further refining prompts and enhancing viewpoint diversity in future iterations.

### D. *Evaluation on Dialogue Transformation through LLM Feedback*

We evaluated RQ3: whether the feedback generated by our framework leads to desirable changes in parent–child dialogue or behavior. To simulate post-feedback conversations, we input both $F_{final,child}$ and $F_{final,adult}$ into the framework along with the original dialogue data $D$, and generated follow-up dialogues. We interpreted these dialogues as proxies for behavioral change and assessed them accordingly.

Three human raters evaluated the results on eight criteria: four child-focused dimensions, Emotional release (C-1), Self-esteem (C-2), Autonomy (C-3), and Constructive dialogue (C-4); and four parent-focused dimensions, Empathic tone (P-1), Bias awareness (P-2), Specificity of support (P-3), and Overall transformation (P-4). Each dimension was rated on a five-point Likert scale.

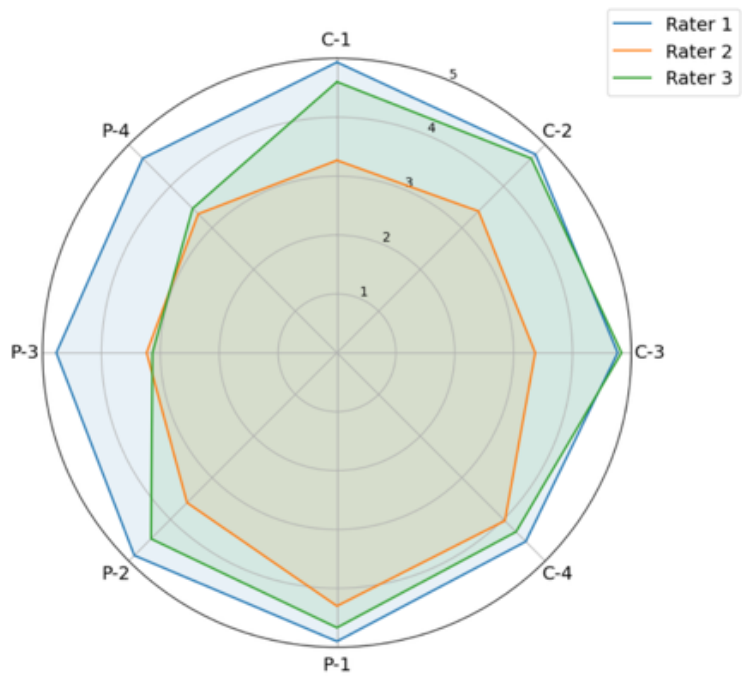

Fig. 7. Evaluation of Simulated Post-Feedback Dialogue.

Fig. 7 presents the evaluation results. Rater 1 and Rater 3 assigned consistently high scores, particularly in C-1, P-1 and P-2, where scores averaged above 4.5. Rater 2 gave slightly more reserved ratings, especially in C-3 and C-4, but still generally scored above 3.0. Notably, C-4 received a score of 4.0 or higher from all raters, indicating that post-feedback dialogues demonstrated clearer emotional expression and more constructive parent–child exchanges. In P-2, the raters showed varied scores, reflecting differences in how clearly the system's feedback conveyed parental bias awareness.

These results suggest that a role-playing LLM-based multi-agent feedback has the potential to foster positive changes in both content and tone of parent–child interactions.

## VI. Conclusion

This study developed a role-playing LLM-based multi-agent support framework that addresses ideal parent bias and suppressed emotion within families. Scenario-based experiments demonstrated the framework's ability to detect suppressed emotion, generate empathetic and actionable feedback for both children and parents, and support positive transformation in dialogue.

However, the framework has not yet been evaluated in real family settings or long-term use, due to the difficulty of such testing. Future work should focus on deployment in real-world contexts, confidence calibration, and more diverse perspectives. Broader applications beyond families may also extend the reach of emotionally supportive AI. Ethical considerations, including privacy protection and avoiding over-reliance on AI feedback, must also be addressed to ensure safe and responsible use.

## Acknowledgment

This work was supported in part by the Nippon Foundation HUMAI Program.